# Magnetic resonance spectroscopy of single centers in silicon quantum wells

Nikolay T. Bagraev,[*] Leonid E. Klyachkin, Andrey A. Kudryavtsev, Anna M. Malyarenko

Ioffe Physical-Technical Institute, 194021, St.Petersburg, Russia

**Abstract**

We present the new optically-detected magnetic resonance (ODMR) technique which reveals single point defects in silicon quantum wells embedded in microcavities within frameworks of the excitonic normal-mode coupling (NMC) without the external cavity and the hf source.

[*] Corresponding author. Tel.: +7-812-292-73-15; fax: +7-812-297-10-17; e-mail: impurity.dipole@mail.ioffe.ru.

# 1. Introduction

The diffusion of boron is known to be controlled by means of adjusting the fluxes of self-interstitials and vacancies thereby forming the p-type silicon quantum wells (Si-QW) confined by the δ-barriers heavily doped with boron on the n-type Si (100) surface (Fig. 1a) [1]. The boron centers inside the δ-barriers are found to be the impurity dipoles, $B^+$-$B^-$, which are a basis of their high temperature superconductor (HTS) properties, $T_c$=145 K, $H_{c2}$=0.22 T, if the sheet density of the 2D holes in the Si-QW becomes to be more than $10^{11}$ cm$^{-2}$ [2, 3]. The superconductor gap appeared to result in the THz and GHz generation under applied voltage [2, 3]. Spectroscopic studies confirmed this pattern and furthermore showed that the hf emission can be enhanced, if the Si-QW is incorporated into the fractal silicon microcavity system which is created between self-assembled microdefects induced by the same fluxes of self-interstitials (Fig. 1b) [1]. The goal of the present work is to use the excitonic normal-mode coupling (NMC) with a single Si-QW inside a 1λ silicon microcavity to study the ODMR of single point defects inserted into the Si-QW in the absence of the external magnetic field, the external cavity resonator and the hf source. The excitonic NMC regime appears to favour the giant triplet-singlet splitting caused by the exchange interaction of the impurity electron states with the s-p electronic states of the host Si-QW which is revealed by the transmission spectra under the formation of the bound exciton at an impurity center [4, 5].

# 2. Methods

The n- type Si (100) wafers were oxidized at 1150°C in dry oxygen containing $CCl_4$ vapors. Short-time boron doping was done from the gas phase under fine surface injection of both self-interstitials and vacancies into the window, which was cut in the oxide overlayer after preparing a mask and performing the subsequent photolithography.

The four-point probe under layer-by-layer etching and SIMS measurements allowed the depth of the $p^+$-diffusion profile, 8 nm. The CR angular dependencies have shown that the profile prepared contains the p-type Si-QW confined by the δ-barriers heavily doped with boron (Fig. 1a) [1]. The energy positions of the two-dimensional subbands of holes in the Si-QW and the value of the superconductor gap, $2\Delta=0.044$ eV, caused by the superconductor δ-barriers were determined in the studies of the far-infrared and tunneling spectroscopy (Figs. 2a and 2b) [2, 3, 6]. The superconductor gap appeared to be the source of the THz emission due to the Josephson transitions self-assembled in the sandwich structure (Figs. 1a, 3a and 3b). This THz emission can be enhanced by varying the geometrical dimensions of the δ-barrier, which define the THz cavity embedded in the Si-QW plane. Besides, the STM images have shown that this single Si-QW confined by the HTS δ-barriers is incorporated into the microcavity system of the fractal type formed by the microdefects of the self-interstitials type (Fig. 1b). The NMC regime of this fractal microcavity system with the single Si-QW appeared to result in the enhancement of both infrared absorption and photoluminescence in the spectral range of the Rabi splitting, which is caused by the bound excitons at impurity centers [4, 5].

3. Results

Two silicon microcavities are revealed by the angular resolved transmission spectra that exhibit the excitonic NMC regime with a single Si-QW in the spectral range of the Rabi splitting at T=300 K (Figs. 4a and 6a) [7, 8]. The angular resolved measurements have revealed the strong coupling regime by an anti-crossing behavior between polariton states in the microcavity embedded in the Si-QW containing the carbon-hydrogen acceptor centers [9]. The NMC regime is found to give rise to the enhancement of bound exciton absorption (Figs. 4a, 4b and 5) and photoluminescence (Figs. 6a, 6b and 7) in the spectral range of the Rabi splitting. Moreover, the exciton localization at the carbon-hydrogen acceptor centers appeared to cause the giant triplet-singlet splitting in the absence of the external magnetic field which is created by strong coupling of the impurity states with the s-p electronic states of the host Si-QW (Figs. 4a and 6a). This strong sp-impurity states mixing is revealed by the angular resolved absorption and photoluminescence that seem to reveal the ODMR spectra in zero magnetic field under the NMC conditions (Figs. 4b and 6b), because the EPR frequency is able to be selected from the THz range generated by the δ-barriers confining the Si-QW being in self-agreement with the splitting of the triplet sublevels in the exchange field induced by the bound exciton (Figs. 5 and 7) [4, 5].

Two different carbon-hydrogen acceptor centers, $C_s$-$C_i$-H [9], have been identified by measuring the ODMR spectra. The first center that is sensitive to the frequency value of 0.12 THz seems to be associated with the presence of the $E_c$ - 0.37 eV acceptor level which appears to give rise to the M-line 760.8 meV photoluminescence (Figs. 4a and 5) [10]. The bound exciton absorption is predominant in the ODMR spectrum, because the energy of optical transition from the HH1 subband of 2D holes to the $E_c$ - 0.37 eV acceptor level appears to be in resonance with the HH3-HH1 optical transition in the Si-QW. It should be noted that the double quantum transitions seem to be exhibited in the ODMR spectrum (Fig. 4a). The second center that is sensitive to the frequency value of 0.087 THz seems to be associated with the presence of the $E_c$ - 0.2 eV acceptor level which appears to give rise to the H-line 925.6 meV photoluminescence (Figs. 6a, 6b and 7) [11]. The phonon replica are observed in the photoluminescence spectrum that reveals also the giant triplet-singlet splitting, 5 meV. The multiquantum transitions are seen to contribute to the ODMR spectrum (Fig. 6b).

Finally, the frequency values of 0.12 THz and 0.087 THz revealed by the modulation of the transmission spectra correlate with the splitting of the triplet sublevels in the exchange field induced by the bound exciton which appears to result from the ODMR spectra.

## 4. Summary

The excitonic normal-mode coupling (NMC) with the single p-type Si-QW incorporated in the 1λ silicon microcavity on the n-type Si (100) wafer has been identified at T=300 K in the studies of the transmission spectra which have revealed the ODMR of the single carbon-oxygen acceptor center in the absence of the external magnetic field, the external cavity and the hf source.

## 5. Acknowledgement

The work was supported by the SNSF programme (grant IB7320-110970/1), RAS-QM (grant P-03. 4.1).

**References**

1. N.T. Bagraev et al., Defect and Diffusion Forum **237-240** (2005) 1049.

2. N.T. Bagraev et al., Physica C **437-438** (2006) 21.

3. N.T. Bagraev et al., Physica C **468** (2008) 840.

4. N.T. Bagraev et al., Physica B **340-342** (2003) 1074.

5. N.T. Bagraev et al., Physica B **340-342** (2003) 1078.

6. N.T. Bagraev et al., J. Phys.: Condens. Matter, **20** (2008) 164202.

7. R. Houdre et al., Phys.Rev.B **49** (1994) 16761.

8. G. Khitrova et al., Rev. Mod. Phys. **71** (1999) 1591.

9. A. Mainwood, Mat. Sci. Forum, **258** (1997) 253.

10. A.N. Safonov et al., Mat. Sci. Forum, **258** (1997) 259.

11. A.N. Safonov et al., Mat. Sci. Forum, **258** (1997) 617.

**Captions**

Fig. 1. (a) The ODMR transmission experiment with the p-type silicon quantum well confined by the δ-barriers containing the dipole centers of boron, which is prepared on the n-type Si (100) surface. The effective exchange field is caused by forming the bound excitons under the optical illumination with linearly polarized light. (b) The model of a fractal microcavity system.

Fig. 2. Electroluminescence spectrum (a) that defines the energies of the two-dimensional subbands of 2D holes in the p-type Si-QW confined by the HTS δ-barriers (b).

Fig. 3. Transmission spectra that exhibit the 0.12 THz (a) and 0.087 THz (b) modulation due to the Josephson transition control of the optical transitions between the subbands of 2D holes in the Si-QW that is incorporated in corresponding 1λ microcavity.

Fig. 4. (a) Spectral dependences of the light transmission coefficient that demonstrates at T=300K the excitonic normal-mode coupling with the Si-QW containing the carbon-hydrogen acceptor center being incorporated in the 1λ microcavity. (b) The giant triplet-singlet splitting revealed by the NMC regime allows the ODMR spectrum (0.12 THz) by the triplet sublevel absorption of the bound exciton at the carbon-hydrogen acceptor center.

Fig. 5. The one-electron band scheme of the p-type Si-QW confined by the HTS δ-barriers that contains the carbon-hydrogen acceptor center. The bound exciton created at this center under optical pumping results in the giant triplet-singlet splitting. The presence of the $E_c$ - 0.37 eV acceptor level appears to be associated with the M-line 760.8 meV photoluminescence [10].

Fig. 6. (a) Spectral dependences of the light transmission coefficient that demonstrates at T=300K the excitonic normal-mode coupling with the Si-QW containing the carbon-hydrogen acceptor center being incorporated in the 1λ microcavity. (b) The giant triplet-singlet splitting revealed by the NMC regime allows the ODMR spectrum (0.087 THz) by the triplet sublevel photoluminescence of the bound exciton at the carbon-hydrogen acceptor center.

Fig. 7. The one-electron band scheme of the p-type Si-QW confined by the HTS δ-barriers that contains the carbon-hydrogen acceptor center. The bound exciton created at this center under optical pumping results in the giant triplet-singlet splitting. The presence of the $E_c$ - 0.2 eV acceptor level appears to be associated with the H-line 925.6 meV photoluminescence [11].

Fig. 1.

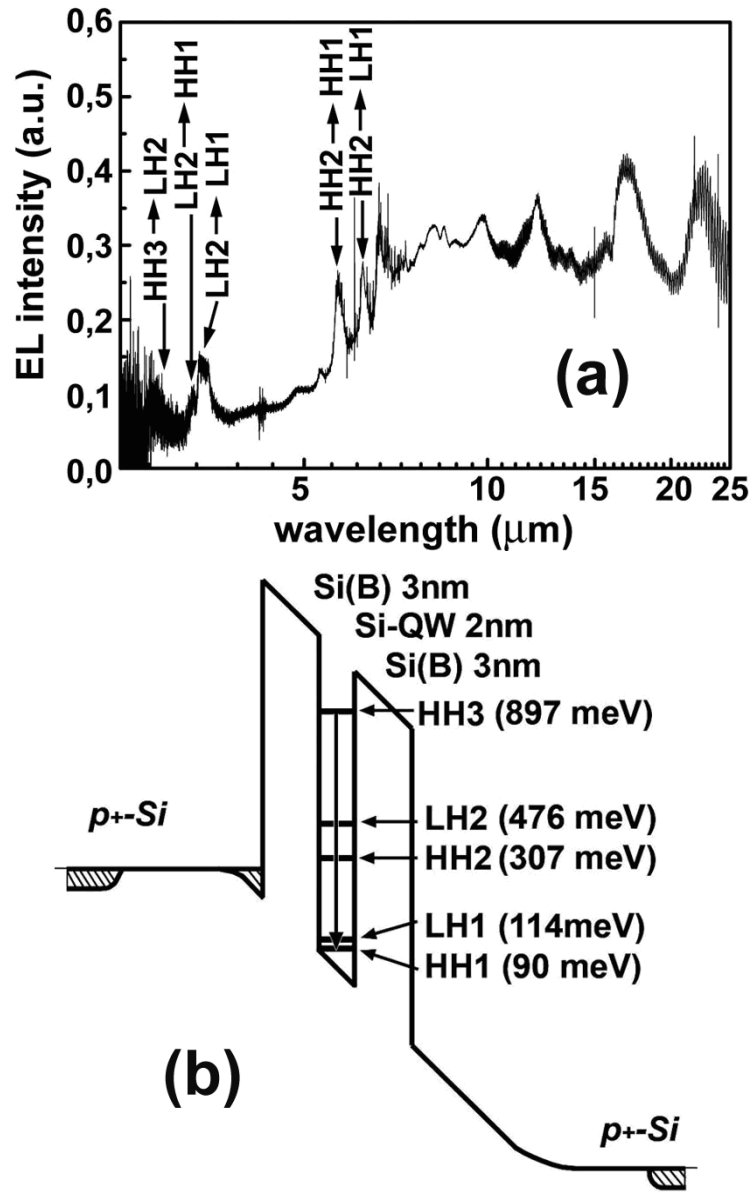

Fig. 2.

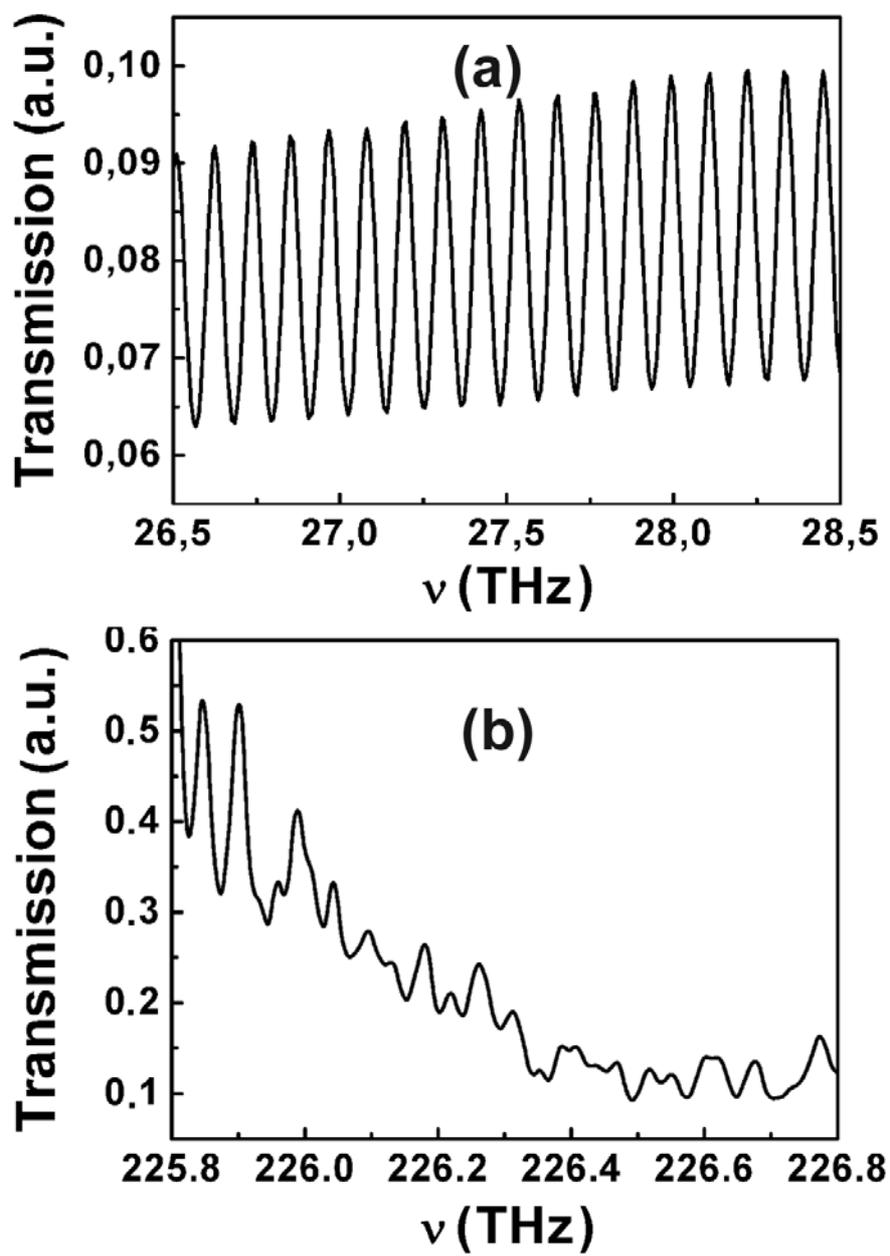

Fig. 3.

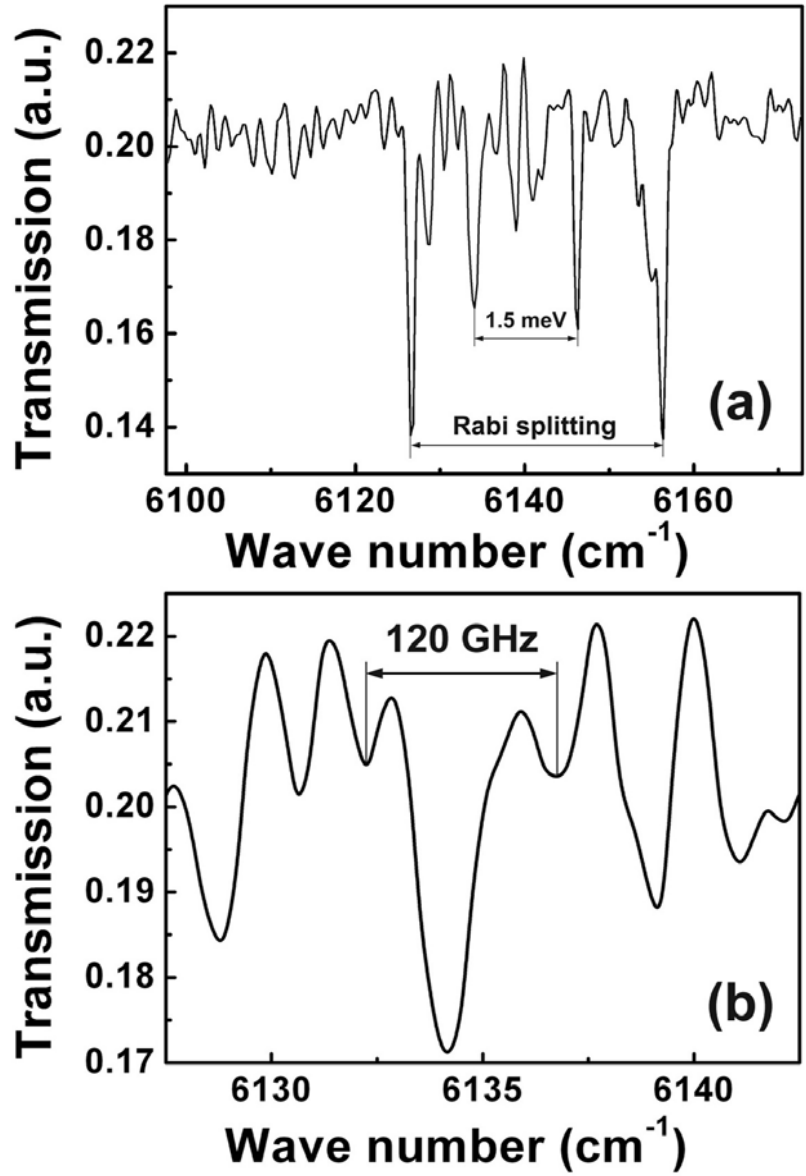

Fig. 4.

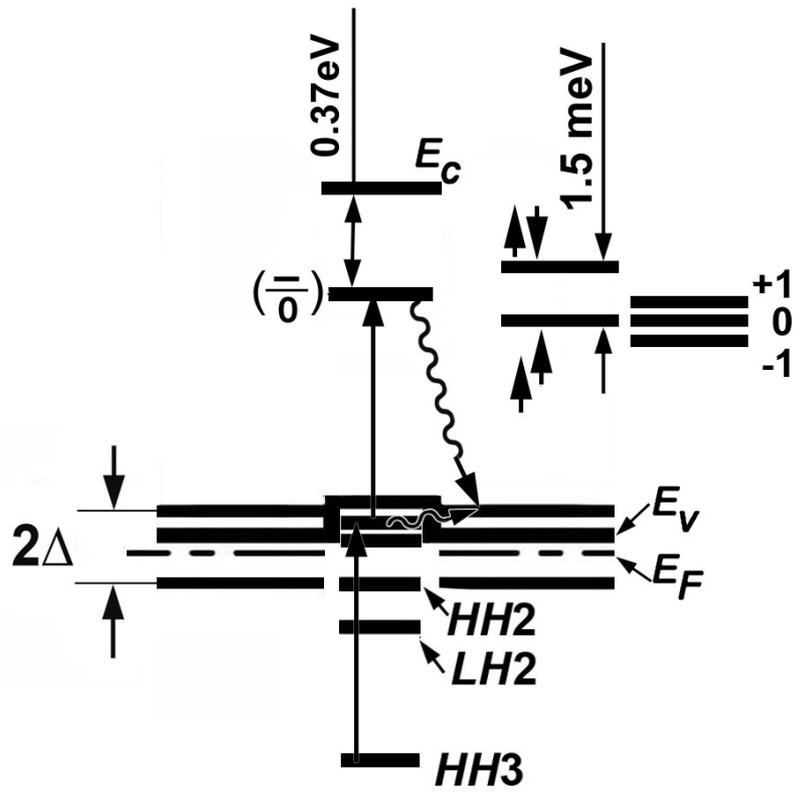

Fig. 5.

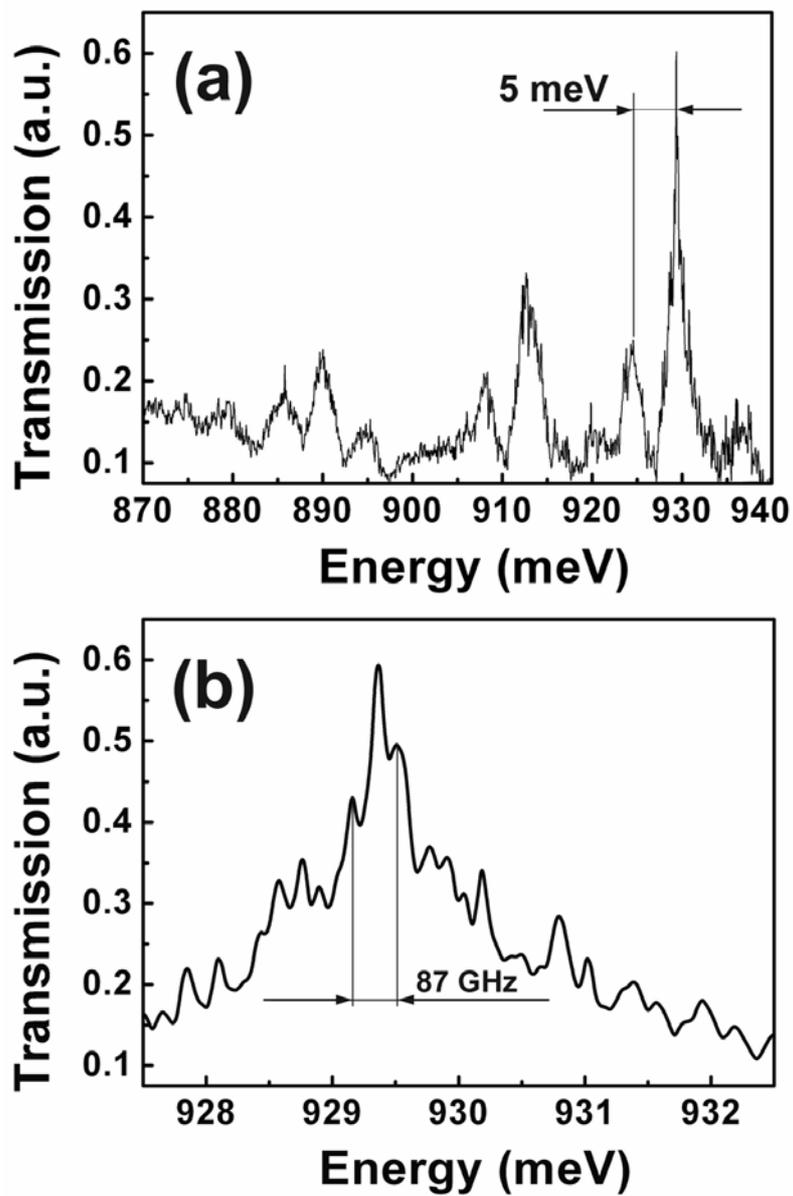

Fig. 6.

Fig. 7.